# Opto-mechanical designs for the HARMONI Adaptive Optics systems

Kjetil Dohlen[*a], Timothy Morris[b], Javier Piqueras Lopez[c],
Ariadna Calcines-Rosario[b], Anne Costille[a], Marc Dubbeldam[b], Kacem El Hadi[a], Thierry Fusco[a,d],
Marc Llored[a], Benoit Neichel[a], Sandrine Pascal[a], Jean-Francois Sauvage[a,d], Pascal Vola[a],
Fraser Clarke[e], Hermine Schnetler[f], Ian Bryson[f], Niranjan Thatte[e]

[a] Aix Marseille Univ, CNRS, CNES, LAM, Marseille, France
[b] Durham University, Centre for Advanced Instrumentation, Durham DH1 3LE, United Kingdom
[c] Centro de Astrobiologica (CSIC/INTA), 28850 Torrejón de Ardoz, Madrid, Spain
[d] ONERA, B.P.72, 92322 Chatillon, France
[e] Department of Physic, University of Oxford, Oxford OX1 3RH, United Kingdom
[f] UKATC, Royal Observatory Edinburgh Blackford Hill, Edinburgh EH9 3HJ, United Kingdom

**ABSTRACT**

HARMONI is a visible and near-infrared integral field spectrograph equipped with two complementary adaptive optics systems, fully integrated within the instrument. A Single Conjugate AO (SCAO) system offers high performance for a limited sky coverage and a Laser Tomographic AO (LTAO) system provides AO correction with a very high sky-coverage. While the deformable mirror performing real-time correction of the atmospheric disturbances is located within the telescope itself, the instrument contains a suite of state-of-the-art and innovative wavefront sensor systems. Laser guide star sensors (LGSS) are located at the entrance of the instrument and fed by a dichroic beam splitter, while the various natural guide star sensors for LTAO and SCAO are located close to the science focal plane. We present opto-mechanical architecture and design at PDR level for these wavefront sensor systems.

## 1. INTRODUCTION

HARMONI [1] is a visible and near-infrared integral field spectrograph providing the ELT's core spectroscopic capability. It will exploit the ELT's scientific niche in its early years, starting at first light. To get the full sensitivity and spatial resolution gain, HARMONI will work at diffraction limited scales. This will be possible thanks to two complementary adaptive optics systems [2], fully integrated within HARMONI. The first is a Single Conjugate AO (SCAO) system offering high performance for a limited sky coverage. The second is a Laser Tomographic AO (LTAO) system, providing AO correction with a very high sky-coverage. Both AO modes for HARMONI have gone through Preliminary Design Review at the end of 2017, and enter their Final Design phase starting in 2018 ending in 2020.

While the deformable mirror performing real-time correction of the atmospheric disturbances is located within the telescope itself, instruments are in charge of providing the wavefront measurements controlling this correction. To this end, HARMONI contains a suite of state-of-the-art and innovative wavefront sensor systems. These are distributed within the instrument according to their various functions. The laser guide star sensors (LGSS) are located at the entrance of the instrument and fed by a dichroic beam splitter reflecting the sodium line. The various natural guide star sensors (NGSS) are located close to the science focal plane. LTAO requires natural guide stars for fast sensing of tip-tilt and focus and for disentangling the effects of slowly varying non-common path errors. The NGSS also contains the SCAO wavefront sensor, consisting of a high-order pyramid wavefront sensor aided by a slow, low-order Shack-Hartmann sensor and a low-order deformable mirror.

In this paper, we present the results of the PDR design studies for these various wavefront sensor systems, including system studies and opto-mechanical architecture and design, with particular attention given to some of the more original aspects of the designs.

---

[*] E-mail: kjetil.dohlen@lam.fr

**1.1. Instrument Product Break-down and interfaces**

The HARMONI instrument (Figure 1) is built up of four main hardware systems. The integral field spectrograph (IFS) [3] consists of a 4m high and 3.25m diameter cylindrical cryostat vessel sitting on a rotator, housing scale-changing optics, image slicer units and specrograph modules. The calibration and relay system (CARS) consists of a relay system (FPRS) in the form of a cooled Offner relay producing a vertical optical axis and a focal plane at the entrance of the IFS, a calibration unit, and the instrument structure (ISS) allowing to present these elements to the telescope optical axis, 6m above the Nasmyth platform. The two remaining systems are the wavefront sensor systems, one working on artificial laser guide stars (LGSS), and the other working on natural guide stars (NGSS).

While the LGSS is supported by the ISS, interfacing optically via a dichroic beam splitter located 1m downstream of the telescope focus, the NGSS is mounted on top of the IFS cryostat and rotates together with the science field. This allows for the alignment between science and natural star sensors to be maintained within a resolution element. The NGSS interfaces optically with both CARS and IFS by operating around the vertical optical axis, picking off natural guide stars located within the 2 arcminute-diameter patrol field provided by the relay system and transmitting the 12 arcminutes diameter science field. To minimize thermal background radiation, the NGSS volume is cooled to 20 degrees below ambient, sharing a cold, dry atmosphere with the relay system through a rotating joint.

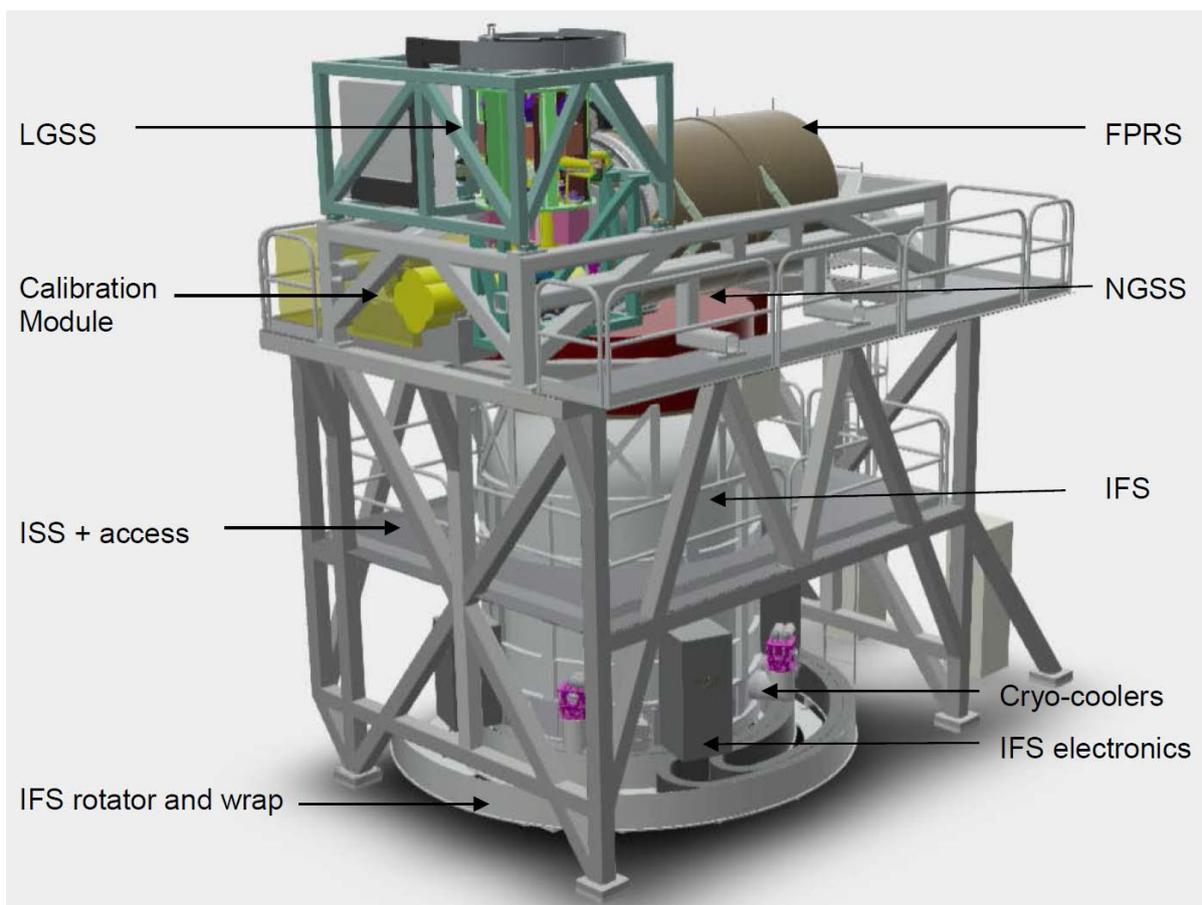

Figure 1. 3D rendering of the HARMONI instrument indicating major systems and components.

## 2. SINGLE CONJUGATE ADAPTIVE OPTICS

The SCAO sub-system of HARMONI is located within the NGSS system. It consists of an optical bench containing a deployable dichroic beam splitter, object selection and beam control opto-mechanics, and the wavefront sensor itself, based on the pyramid principle. The dichroic beam splitter intercepts the scientific beam, reflecting the visible light while transmitting the infrared beam towards the IFS. It covers a field of 30" diameter, allowing to select guide stars within or around the scientific field. The size of the SCAO patrol field corresponds to a cross-over point between off-axis SCAO performance and LTAO performance. Three positions are available in the dichroics exchange mechanism, offering two different long-wave cut-off wavelengths (800nm and 1000nm) and a pick-off unit feeding the high-contrast module of HARMONI [4].

The SCAO optical design provides a series of field and pupil images in which active devices are placed as shown in Figure 2. First, a mechanism places a dichroic plate into the science beam as it converges towards the focal plane at the entrance of the IFS, projecting a conjugate image plane up and out of the beam. A concave field mirror projects a pupil image onto the Object Selection Mirror (OSM), a high precision, large stroke tip-tilt mechanism, allowing navigation within the SCAO patrol field in order to select the required guide object. A couple of doublet lenses produces a second pupil plane. Preceded by a second dichroic, photons red wards of 700nm are transmitted on via a low-order deformable mirror (LODM), while "blue" photons in the band 650 to 700nm are sent to a low-order Shack-Hartmann sensor, the BlueSH. These two elements constitute the Low-Order Module which provides stabilization both of object pointing and low-order wavefront errors as will be described further on.

Next, the beam Correction Module contains a derotator, ensuring stabilization of the telescope pupil on the wavefront sensor camera, and an atmospheric dispersion corrector (ADC). While the science path does not require such correction thanks to its spectroscopic imaging capability, it is essential for diffraction limited operation of the pyramid wavefront sensor. The module also contains a pupil correction unit, allowing lateral and dimensional pupil stabilisation. This includes a pupil zoom, as will be described further on. The final module contains the pyramid wavefront sensor [5] based on a modulated achromatic double pyramid design [6].

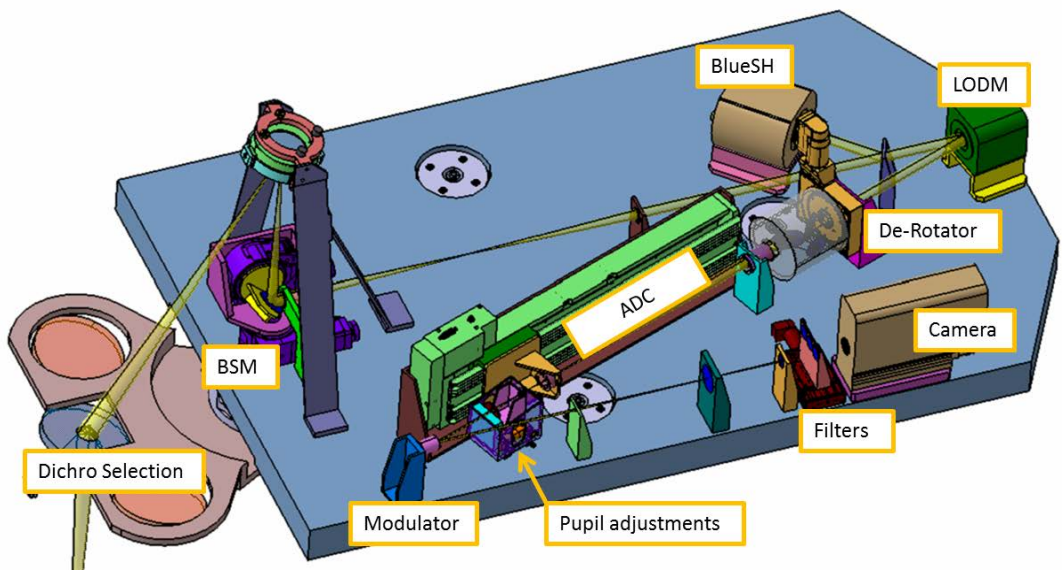

Figure 2. Identification of active devices in the SCAO subsystem.

While most of these functions are quite classical, two of them, the Low-Order Module and the Pupil Zoom, are original to our system. We describe these in more detail in the following.

**2.1. Low-Order Module**
This module offers a slow (<0.1 Hz), low-order wavefront correction loop consisting of a 13x13 sub-aperture Shack-Hartmann sensor and corresponding resolution deformable mirror. Photons in the band lambda 650 to 700nm, which are of little use for the pyramid sensor because of the increasingly large residual phase errors, are reflected off a fixed dichroic plate located just upstream of a pupil image, and aliment this sensor which operates at a low temporal bandwidth. It measures the slopes of the wavefront at that point, comparing them to reference slopes obtained during a calibration procedure. The slope errors are projected onto a phase map which is transmitted to the LODM. While this might appear to constitute an open loop operation, it actually represents closed loop operation since it enters into the full SCAO loop, as seen in Figure 3: any wavefront deformations seen by the pyramid sensor will immediately be corrected by the M4 adaptive mirror in the telescope. Since the BlueSH has a large dynamic range, matched by the large dynamic range of the LODM, large low-order non-common path errors will be absorbed here, allowing the pyramid sensor, with its limited dynamic range, to work within its linear regime close to zero. This concept is analogous to the differential tip-tilt loop in SPHERE.

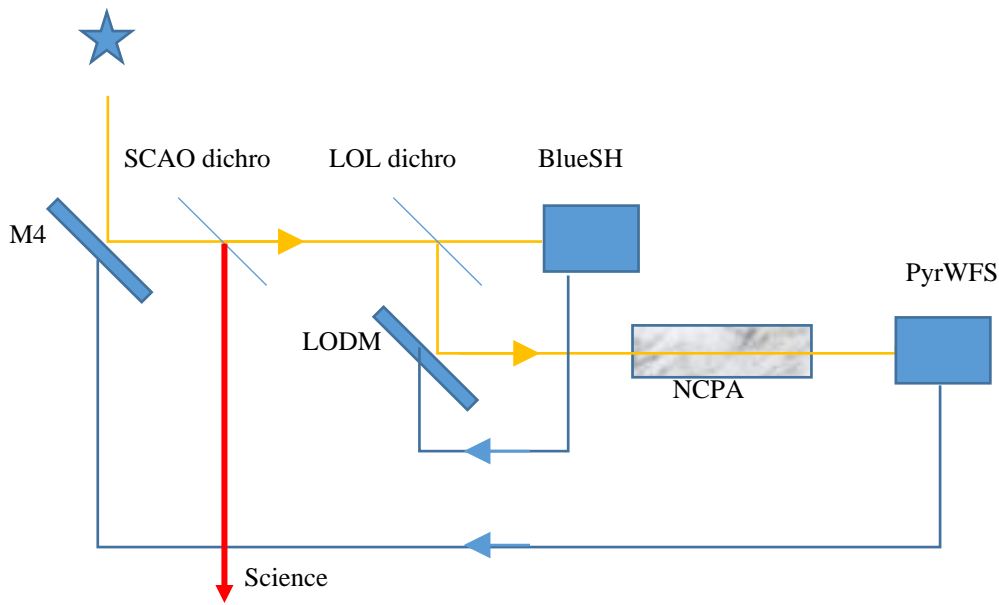

Figure 3. Illustration of the Low-Order Loop, BlueSH + LODM, working in closed loop thanks to its position within the main AO loop, PyrWFS + M4.

This loop ensures several critical functions:

- **Absolute pointing:** The loop acts as a differential tip-tilt sensor, fixing the science field position with respect to the known position of the guide star. In particular, any wobble movements due to imperfections in de-rotator or ADC alignment, as well as thermomechanical movements within the pyramid sensor system are taken out. Dynamically updating the reference slopes will also allow removing differential atmospheric dispersion in order to stabilize the infrared image. Since the optical bandwidth is narrow, this correction can be based only on knowledge of pointing and atmospheric parameters (temperature, pressure, humidity) without prior knowledge or assumptions about stellar type. With a diffraction limited resolution of the BlueSH images 13 times larger than that of the telescope itself, i.e. at lambda=650nm, 45 mas and assuming centroiding to 1/100 resolution element, this allows pointing stabilization to 0.45 mas.
- **NCPA compensation:** Non-common path aberrations can be calibrated and compensated by modification of the reference slopes. This includes low-order aberrations in the science path, in particular the astigmatism introduced by the SCAO dichroic plate, but also, low-order aberrations in the down-stream SCAO. This reduces pressure on optical aberration correction for the down-stream optics, allowing for design simplifications.
- **In High-Contrast mode**, reference slopes can be modified dynamically in response to online wavefront measurements provided by the Zernike sensor [7] foreseen in the High-Contrast Module [4].

## 2.2. Pupil Zoom

The use of a field selector based on a concave field mirror projecting a pupil image onto a high precision, large stroke tip-tilt mechanism provides for a conceptually elegant and operationally simple system, contrary to the classical periscope-type field selectors. It does introduce a slight pupil distortion, however, reaching a 10% variation in pupil diameter, see Figure 5. A pupil zoom system [8] compensates for this distortion, ensuring constant distance between an image plane and a pupil plane, while changing the focal length, hence pupil magnification. The length of the system is ~80mm for a median focal length of 100mm. The two outer lenses move together by 2 mm, while the middle lens moves faster, covering a distance of 5mm, see Table 1. Dimensional control of pupil diameter is required to a precision of <0.03subaperture, ie 3e-4 pupil diameter. This requires a movement with a precision of about 5 microns. Adding lateral motions on one of these lenses allows for lateral pupil position adjustment as well.

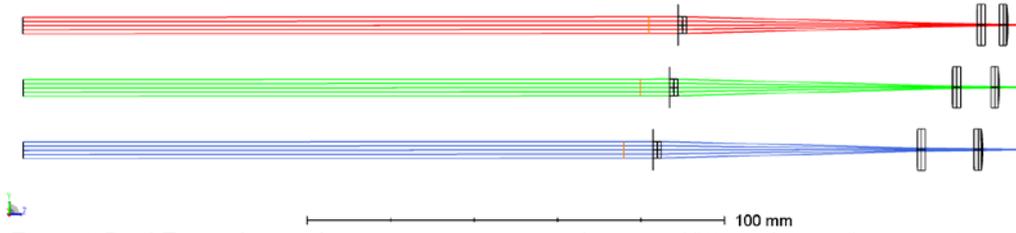

Figure 4. Pupil Zoom, showing lens movements corresponding to +/-10% pupil magnification variations.

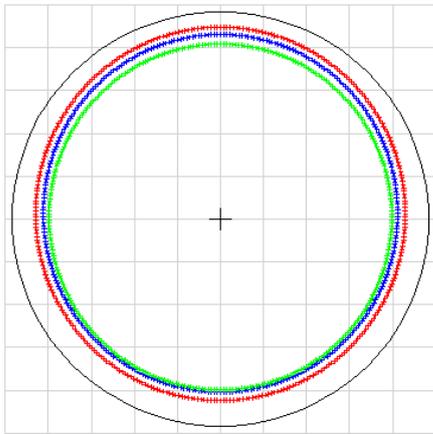

Figure 5. Pupil image for different guide-star positions.

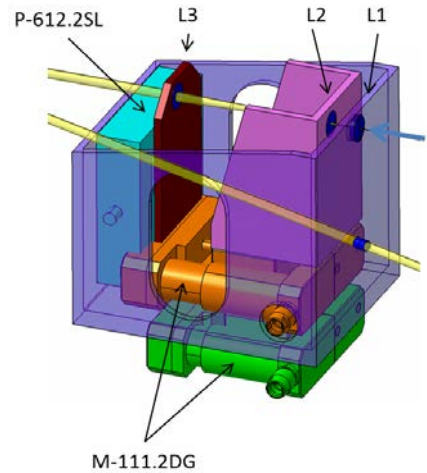

Figure 6. 3D implementation of the Pupil zoom and TT module. It consists of two stacked linear stages for axial movement and an X-Y stage for lateral movement of L3.

Table 1. Position of the three lenses ensuring constant pupil and image positions for a range of 8% pupil size variation.

| Image | L_1 | L_2 | L_3 | Pupil | *f* |
|---|---|---|---|---|---|
| 0 | 1.546 | 8.737 | 81.546 | 240 | *95.65* |
| 0 | 2.540 | 11.556 | 82.540 | 240 | *100* |
| 0 | 3.471 | 13.840 | 83.471 | 240 | *103.31* |
|  | *2mm* | *5mm* | *2mm* |  |  |

# 3. LASER GUIDE-STAR SENSORS

The six ELT laser guide stars are analysed using the LGSS system. Located at the entrance of HARMONI, it receives the laser light reflected off a large dichroic beam splitter. A fold mirror sends the 6 laser beams vertically upwards where they are individually intercepted by the six wavefront sensor modules. These are mounted in a rotating core structure, controlled such as to maintain the pupil stabilized within the wavefront sensors.

The wavefront sensors are Shack-Hartmann sensors with 80 sub-apertures across the telescope pupil. The optical train is composed of pupil imaging optics and a pupil plane housing a beam steering mirror, to which we will return further on. Following this, camera optics creates a telecentric focus, where the focal ratio stays constant regardless of object distance (ie sodium layer height). A trombone system composed of a series of flat mirrors, intercepts this beam, allowing modification of the optical path and providing a fixed focus position. A pupil tip-tilt mirror located in this focus allows fine control of the pupil position, and a pupil imaging lens projects the pupil onto the microlens array.

## 3.1. Optical design: the optimal laser asterism

The optimal asterism radius varies according to apparent height of the sodium (Na) layer, itself a function of Zenith angle, see Figure 7. This can be understood by considering the cartoons showing how the cone effect allows for more or less efficient sampling of the turbulent atmosphere. The optimal asterism in the case of a small scientific field is clearly the one which corresponds to the localization of the laser stars at the telescope's rim ray. Consequently, as the telescope points further off Zenith, the optimal asterism radius gets smaller.

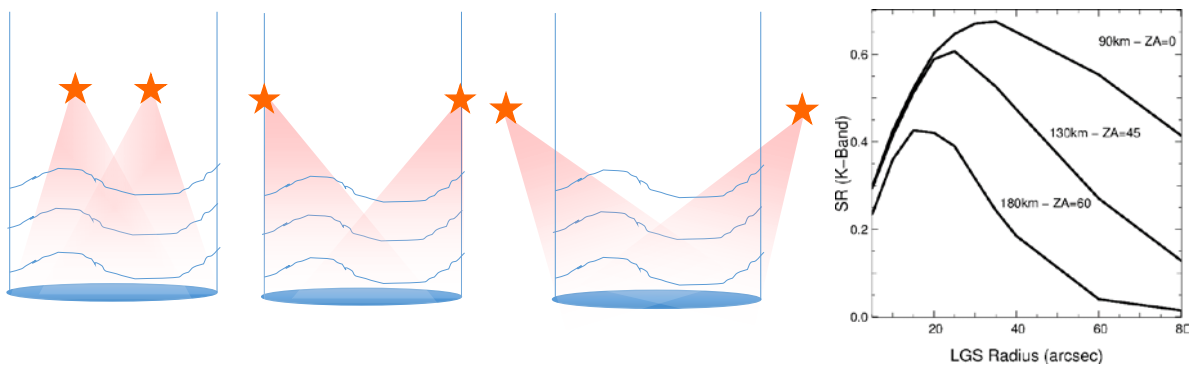

Figure 7. Left: Illustration of various asterism radii. The second cartoon corresponds to optimal asterism. Right: Strehl ratio as function of LGS asterism radius for three different Zenith angles (ZA).

Our wavefront sensor design accounts for this evolution of the asterism radius by allowing for an adjustment of the sensor's pointing direction using the beam-steering mirror. The most critical part of the optical design for these sensors are the pupil imaging lenses at the entrance. They need to produce well-corrected images at vastly different object distances (the laser focus moves by several meters when going from Zenith to 30 degrees from the horizon), and they must tolerate the adjustment in asterism radius. An appropriate design has been found using two free-form lenses as shown in Figure 8, staying within the required wavefront error budget of 100nm rms and 0.1% pupil distortion.

The optical bench of each WFS module is 1030 mm long and 280 mm wide. The six identical benches will be built as a small series of interchangeable units using a cloning bench, making sure the optical and mechanical interfaces are aligned as required. The interface points in the core structure will be adjusted during initial assembly so that in case of intervention requiring dismounting one of the benches, a spare bench can be installed with no need for optical alignment on the Nasmyth platform. This strategy orients the design of this assembly, both in terms of mounting features and in terms of alignment methodology.

It is noticeable that this laser wavefront sensor design, thanks to its modularity, can be ported to other ELT instrument. This is the case for the MOSAIC instrument which employs four laser wavefront sensors similar to this one [9].

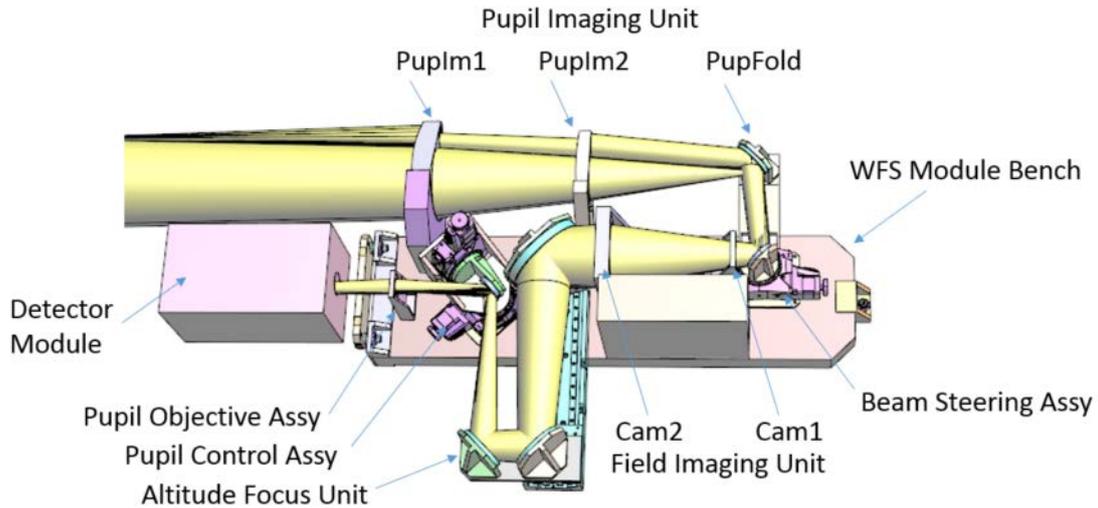

Figure 8. Opto-mechanical design of the LGSS wavefront sensor module.

**3.2. Mechanical structure design**

The LGSS system is composed of a large welded structure, see Figure 9 left, mounted on top of the Harmoni static structure. A rotator and cable wrap are located at the top of this structure and the rotating core structure, Figure 9 right, is hanging off this rotator. This choice minimizes the dimension and hence cost and mass of the rotator, avoiding the need for an optical beam passage through its centre. Steel was chosen as material for both static and rotating structures, avoiding any problems of differential thermal expansion both with respect to the instrument structure, also in steel, and with the steel bearings. These choices will need to be reassessed in the coming phase due to the post-PDR change from steel to aluminium for the instrument structure in order to save mass. A system of flexible blades will probably be required, the difficulty of which being their location at the top of the instrument so they need to be stiff and resistant enough for supporting the large momentums that are induced by horizontal earthquakes accelerations.

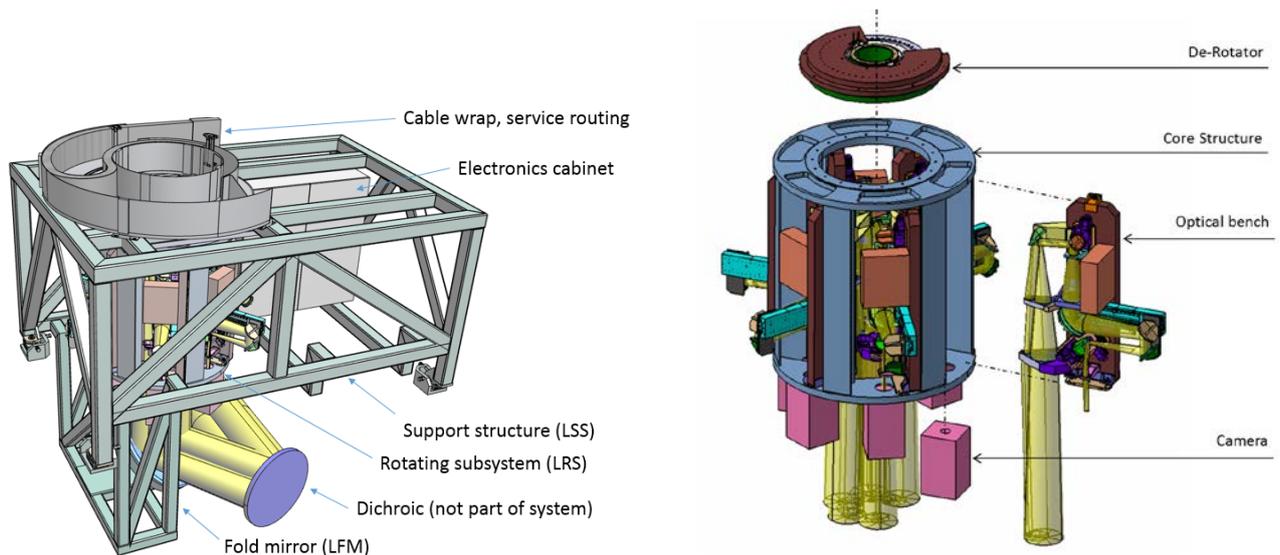

Figure 9. Left: Overall view of the LGSS system. Right: View of the rotating core showing WFS and detector modules.

In the current design, detector modules, containing microlenses, imaging relay, and camera, are separated from the WFS modules. This is due to their size, which is quite consequential. The main factor driving the size is the choice of the LISA detector package provided by ESO [10]. It is itself voluminous, and due to the large separation between the detector chip and the entrance window which does not allow mounting microlenses right in front of the detectors, an optical relay is required. As the needs and design is common with the MAORY LGS WFS [11], it led us to adopt the detector module from their design for this study.

## 4. LTAO NATURAL GUIDE-STAR SENSORS

The LTAO natural guide star sensor (LGNS) module, see Figure 10, provides true wavefront measurements in support of the laser tomographic measurements. This includes a fast (500 Hz), wide field-of-view tip-tilt-focus sensor (TTFS) and a slow (0.1 Hz) Shack-Hartmann 'truth' WFS (TruthS) for measuring slowly varying residual low-order wavefront aberrations left over after laser tomographic (LTAO) correction. It is mounted onto a theta-phi navigation stage, see Figure 11, allowing its pick-off mirror to select any star within a 2 arcminute-diameter technical field. After pickoff, light is split using a dichroic beam splitter between the TruthS observing a wavelength range from 1 - 1.2 μm and the TTFS observing wavelengths from 1.2 - 1.8 μm. Wavelengths below 1 μm are used by a third sensor, also mounted on the same theta-phi stage, the Secondary Guiding Module (SGM) which is used during non-AO observations.

The TTFS consists of a 2x2 SH-WFS system with 8 mas pixel scale covering a 1-arcsecond FOV requiring a low-latency, high speed readout detector of minimum size of 250x250 pixels. In this wavelength range, such performance is achieved using a Saphira-type detector [12].

The TruthS consists of a 10x10 SH-WFS system with 40x40 pixels of 25mas per subaperture. This requires a 400x400 pixel detector that can integrate for up to 10s. Three potential detectors capable of operating at these rates and wavelengths have been identified: SAPHIRA [12], InGaAs [13] and H2RG [14] devices. The InGaAs option is chosen since it is sufficient in terms of performance for this sensor and far more cost and volume efficient than the two other detectors.

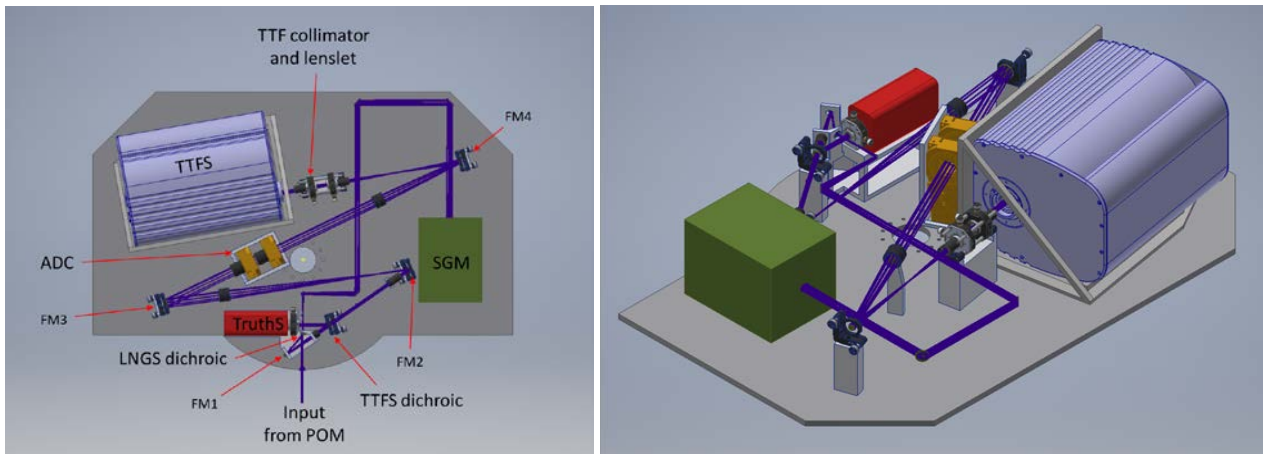

Figure 10. Preliminary mechanical layout of LNGS sensors.

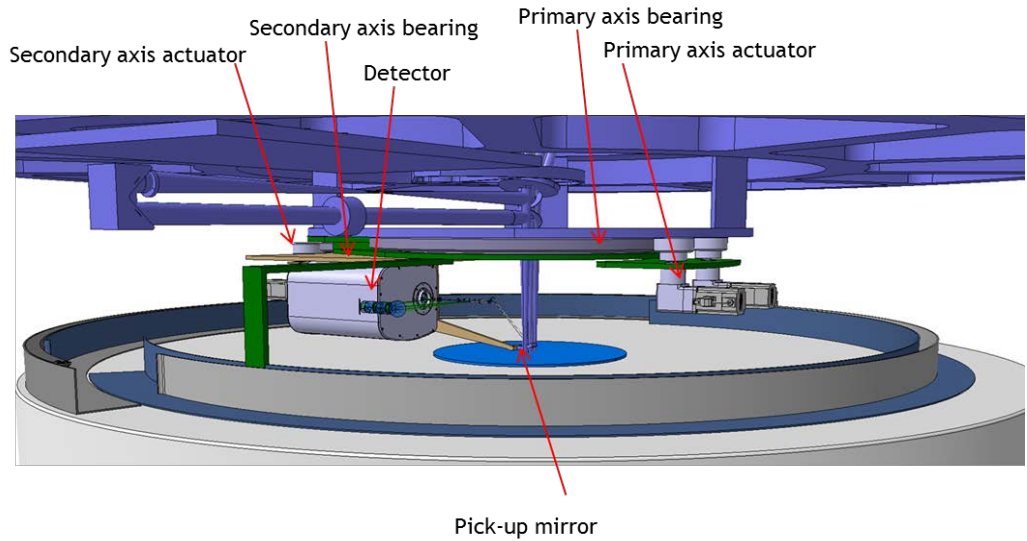
Figure 11. The Low-order wavefront sensor within its theta-phi navigation stage.

### 4.1. Atmospheric Dispersion

Correction of atmospheric dispersion (see Figure 12) is key to the performance of the TTFS, for which concentration of flux within diffraction limited peaks corresponding to 20 meter-class sub-apertures must be optimized. This sensor is therefore equipped with an ADC in the form of two counter-rotating double or triple prisms located in a pupil plane. Detailed design of these prisms, complicated by the need to also correct instrumental dispersion due to the large dichroic located at the entrance of HARMONI required to pick off the laser light, is ongoing. Because of the presence of an ADC, fatally associated with beam wobble, the tip-tilt measured by the TTFS excludes knowledge of absolute pointing at the mas level. Instead, this is left to the TruthS, which operates at much lower speeds and can therefore tolerate operation in the presence of atmospheric refraction.

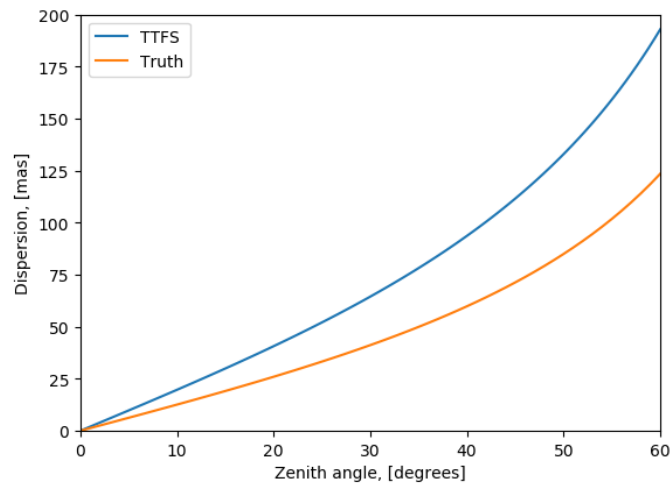
Figure 12. Atmospheric dispersion versus zenith angle across required TTFS and TruthS passbands.

### 4.2. Pupil rotation

While the NGSS system co-rotates with the IFS, hence sees a stable field, the pupil will be seen to rotate at a rate reaching 180 degrees in ~6 minutes at the closest Zenith angle of 1.5°. In case of non-sidereal tracking, which is one of the operating modes of the instrument, further rotation is induced by the theta-phi positioning mechanism, but this additional rotation will be much slower.

For the TTFS, operating at 500Hz with large sub-apertures, this rotation will be handled by taking it into account in the transformation of the measured tip-tilt axes to the telescope-oriented tip-tilt axes. Errors in the knowledge of pupil rotation will translate directly into an error in the applied tip-tilt signal that must be applied to M4/5. Focus measurements are not affected by the pupil rotation.

The situation is rather more complicated for the TruthS, where the larger number of sub-apertures and the stringent (30nm RMS) wavefront error budget allocation increases the sensitivity to rotation errors. Also, the longer integration times introduces smearing of the pupil during exposure. A 3 degree rotation of the wavefront sensor moves the edge of the pupil by just over a quarter of a subaperture, causing errors in all wavefront modes. Even radially symmetric modes can be affected due to aliasing errors. The introduced errors will depend upon the residual aberrations presented to the sensor. There are several sources of these residual errors, including LTAO residuals and field aberrations due to the telescope and HARMONI's own relay. These errors will be concentrated in lower order modes and have an estimated WFE between 460nm and 1061nm RMS depending on NGS location within the field. Here we adopt a 1/3 power law to distribute aberrations across modes, but the result will be highly dependent on this distribution, which knowledge requires a better understanding of residual aberrations within the combined ELT/LTAO system than is currently available. Figure 13 plots the wavefront error caused by a range of pupil rotations in the case of a 460nm RMS residual error and assuming reference slopes to be fixed during exposure. The 30nm RMS budget is reached for wavefront rotation of ~ 2 degrees, implying that for the aberration level and distribution used here, the LGS WFS reference slopes will have to be updated at a rate faster than the TruthS exposure time.

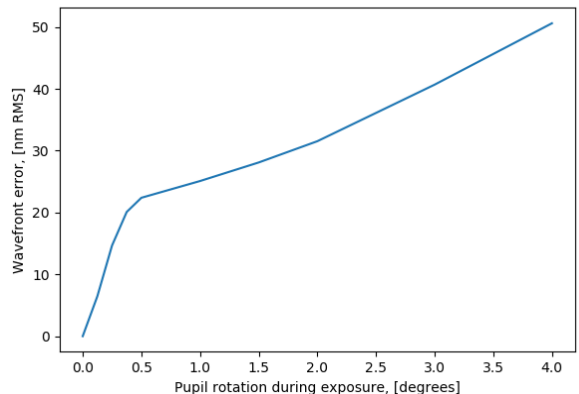

Figure 13 Truth sensor wavefront error due to rotating pupil

## 5. CONCLUSION

We have described the opto-mechanical designs of the adaptive optics wavefront sensors alimenting the two AO modes of HARMONI, LTAO and SCAO. The LTAO mode requires both laser sensors, provided in the LGSS system located at the entrance of the instrument, and natural guide star sensors, provided by the LNGS module within the NGSS system. The NGSS system, mounted on top of Harmoni cryostat and hence co-rotating with the cryostat housing the science instrument, also contains the SCAO sensor. These sensors employ a number of different sensor concepts, ranging from the pyramid sensor used for fast high-order sensing for SCAO, seconded by a low-order Shack-Hartmann (SH) sensor operating together with a low-order deformable mirror forming the low-order loop (LOL), to a collection of six fast, high-order SH sensors used for sensing the laser guide stars in the LTAO system. The natural guide star sensors for the LTAO system employ two low-order SH sensors, a slow 10x10 sub-aperture sensor for truth sensing and a fast 2x2 sensor for tip-tilt and focus sensing.

The unprecedented size of the ELT pupil combined with the requirements in terms of functionality and design volume imposed by the HARMONI instrument led us to consider innovative designs for these sensors. For example, the quest for optimal LTAO performance imposed the use of variable asterism diameter for the laser guide stars required the use of state-of-the-art free-form lenses and beam steering capabilities within each LGS wavefront sensor. The use of a low-order loop in the SCAO sensor is a novel approach to the problem of compensating non-common path aberrations in pyramid-based sensors. Critical performance modelling was required for these sensors, as illustrated by the issues introduced by pupil rotation in the LTAO natural guide-star sensors.

The descriptions given in this paper are representative of the designs made for the preliminary design review, which was passed in December 2017. Work is currently starting to bring this design through the final design within the coming two years, in order to be ready for installation in Chile in 2024.

## ACKNOWLEDGEMENTS
HARMONI is an instrument designed and built by a consortium of British, French and Spanish institutes in collaboration with ESO. JPL acknowledges support from the Spanish Ministerio de Economía y Competitividad through the grant AYA2017-85170-R.

## REFERENCES
[1] Fraser Clarke, Niranjan A. Thatte, et al., "HARMONI @ ELT: status of the AO assisted, first light, visible and near-IR integral field spectrograph at the end of the preliminary design phase." This conference, 10702-62.
[2] Benoit Neichel, et al, "HARMONI at the diffraction limit: from single conjugate to laser tomography adaptive optics." This conference, 10703-39.
[3] Hermine Schnetler, Andy J. Born, Dave J. Melotte, et al. "A novel approach to the development of the HARMONI integral field spectrograph (IFS) using structured systems thinking." This conference, 10705-6.
[4] Alexis Carlotti, François B. Hénault, et al., "System analysis and expected performance of a high contrast module for HARMONI." This conference, 10702-352.
[5] R. Ragazzoni, J. Mod. Opt. 43 (1996) 289.
[6] Jean-Pierre Véran, "Design of the NFIRAOS PWFS module," in *Wave Front Sensing in the ELT era workshop*, Padova, 2017.
[7] M. N'Diaye, K. Dohlen, T. Fusco and B. Paul, "Calibration of quasi-static aberrations in exoplanet direct-imaginginstruments with a Zernike phase-mask sensor," A&A 555, A94 (2013).
[8] Miks and Novak, Applied Optics (2012)
[9] Timothy J. Morris, Alastair G. Basden, Ariadna Calcines-Rosario, et al, "Phase A AO system design and performance for MOSAIC at the ELT." This conference, 10703-43.
[10] Mark Downing, Mark Casali, Enrico Marchetti, Leander Mehrgan, Javier Reyes, "Update on development of WFS cameras at ESO for the ELT." This conference, 10703-69.
[11] Laura Schreiber et al, "The MAORY laser guide star wavefront sensor: design status." This conference, 10703-71.
[12] C-RED1 datasheet, www.firstlight.fr (Dec 2016)
[13] C-RED2 Provisional datasheet, www.firstlight.fr (Feb 2017)
[14] FREDA Camera Technical Requirements Specifications, ESO report ESO-300324, issue 1.7.